\documentstyle[prd,aps,epsf,epsfig]{revtex} 
\begin{document}
\draft
\twocolumn[\hsize\textwidth\columnwidth\hsize\csname@twocolumnfalse\endcsname
%
\title{
Nontrivial stochastic resonance temperature for the
kinetic Ising model}
\author
{Kwan-tai Leung and Zolt\'an N\'eda\cite{NZ}}
\address{
Institute of Physics, Academia Sinica,
Taipei, Taiwan 11529, R.O.C.
}

\maketitle
\centerline{\small (Last revised \today)}

\begin{abstract}
The kinetic Ising model in a weak oscillating magnetic field
is studied in the context of stochastic resonance.  The signal-to-noise
ratio calculated with simulations is found to peak at a nontrivial resonance 
temperature above the equilibrium critical temperature $T_c$.
We argue that its appearance is closely related to the vanishing of
the kinetic coefficient at $T_c$. 
Comparisons with various theoretical results in one and higher 
dimensions are made.

\end{abstract}

\pacs{PACS numbers: 64.60.Ht, 05.40.+j, 05.50.+q}
\vspace{2pc}
]

\vspace{1cm}


\section{Introduction}
A series of recent papers \cite{neda,leung,brey,schimansky,sides}
reveal unequivocally 
the phenomenon of stochastic resonance (SR) \cite{sr} 
in the kinetic Ising model driven by an oscillating magnetic field. 
The possibility of SR was anticipated by
viewing the Ising model as a system of coupled two-state oscillators in a
stochastic force-field which is taken to be thermal fluctuations. In our
previous papers \cite{neda,leung}, we considered the synchronization 
aspects \cite{sync} of SR manifested in the correlation 
function $C(T)$ between 
the magnetic field $h$ and the magnetization $M(T)$ as a function of 
temperature $T$. $C(T)$ was shown to exhibit
two maxima at resonance temperatures $T_{\rm r1}<T_c<T_{\rm r2}$ 
($T_c$ being the critical temperature of the equilibrium system with 
$h=0$). The values of both $T_{\rm r1}$ and $T_{\rm r2}$ 
depend on the driving frequency. 
They converge to $T_c$ in the experimentally relevant,
low frequency limit, thus no
novel characteristic temperature for the kinetic Ising model was observed.
The results are qualitatively the same for two-dimensional (2D) and 
three-dimensional (3D) systems, 
and of course in the one-dimensional (1D) case 
$T_{\rm r1}$ is absent.

Conventionally, SR is characterized by the behavior of 
the signal-to-noise ratio (SNR).
In the case of the 1D kinetic Ising model 
governed by Glauber dynamics\cite{glauber},
the SNR was computed analytically \cite{schimansky}
both for the response of an embedded spin and of the whole
chain, after Glauber's original derivation \cite{glauber}.
Their results (to be discussed in greater detail in Section IV) 
suggest that the SNR exhibits a maximum at a weakly frequency dependent 
resonance temperature $T^{\rm 1D}_r$.
In this manner a new characteristic
temperature for the kinetic Ising model was introduced.

The aim of this work is to study the behavior of the 
SNR for the kinetic Ising model in general spatial dimension $d$.
We are particularly interested in the mechanism responsible for
the maximum of SNR versus temperature, 
its position and the associated scaling behavior 
as a function of system sizes, driving amplitude, driving 
frequency, and dimensionality of the system.

\section {simulation method}

Standard Monte Carlo simulations for the kinetic Ising model  
in an oscillating magnetic field, $h=A \sin(\omega_s t)$, 
on one to four dimensional square-type lattices are carried out.
Heat bath algorithm is used.  
For 1D and 2D, a detailed study of the SNR 
as a function of the driving frequency 
$f_s=2 \pi \omega_s$, driving amplitude $A$, system size
$V=L^d$ and sampling length $N$ are performed. 
Having determined the trend of such dependences,
SNR in 3D and 4D are calculated only for representative 
values of the parameters.
$f_s$ is expressed in units of 1/MCS (inverse Monte Carlo step per site) 
and $A$ in units of the nearest neighbor coupling constant $J$. 
Free boundary conditions are imposed and system sizes up 
to $V=3\times 10^4$ are simulated.

To determine the SNR at each temperature, we follow the time 
evolution of the total magnetization $M(t)$ for $N=2^p$ MCS, 
after sufficient equilibration, 
where $p$ ranges from  $10$ to $12$.
The power spectrum $S(f)$ of $M(t)$ is computed using the Fast 
Fourier Transformation method.
Averaging over 500 to 1000 independent runs, 
the ensemble-averaged power spectrum $<S(f)>$ is obtained. 
A typical $<S(f)>$ is presented in Fig.~1, which shows 
the characteristic sharp peak at the driving frequency $f_s$
along with background noises.  To compute the SNR, 
the noise level near $f_s$ is determined by averaging
$<S(f)>$ over the interval 
$I=[f_s-6/N, f_s-2/N] \bigcup [f_s+2/N, f_s+6/N]$.
The result is denoted by $\mid <S(f)> \mid_I$. 
Taking the height of the peak minus the averaged noise level as the signal, 
we define the SNR in the simulations by:
\begin{equation}
\tilde{R}_{\rm sim}=\frac{<S(f_s)>-\mid <S(f)> \mid_I}{ \mid <S(f)> \mid_I}.
\label{snrsim1}
\end{equation} 

In a typical continuum linear-response calculation of the power spectrum,
one obtains the following general form:
\begin{equation}
S(\omega)=S_0(\omega) + Q \delta(\omega-\omega_s),
\label{spectrum}
\end{equation}
where $S_0(\omega)$ is the zero-field spectrum, and $Q\propto A$ 
is the amplitude of signal.
These two terms correspond to the noisy background and the
sharp peak, respectively, as depicted in Fig.~1.
Conventionally, the SNR is defined by the ratio $Q/S_0(\omega_s)$\cite{sr}.
Thus, in order to compare with theories, some adjustments 
to $\tilde R$ need to be made.
It is straightforward to see that the proper definition 
after replacing $f$ by $\omega$ is
\begin{equation}
R_{\rm sim}={2\pi\over N} \tilde R_{\rm sim},
\label{snrsim2}
\end{equation}
where we have also normalized by the trivial factor $N$ which
arises from discrete Fourier transform so that $R_{\rm sim}$
is independent of $N$.
Hereafter our simulation results will be presented 
in terms of $R_{\rm sim}$.
Notice that in the small frequency and small amplitude limit,
$R_{\rm sim}$ is expected to scale as (cf. \cite{schimansky})
\begin{equation}
R_{\rm sim}(T;V,A) = V A^2 g(T)
\label{snrscale}
\end{equation}
where $g(T)$ is independent of $V$ and $A$.

\section{Simulation results}

In all the dimensions considered, $R_{\rm sim}(T)$ exhibits a
characteristic peak at a resonance temperature $T_r$.
Results for 1D are presented in Figs.~2--3 for the effects of
the driving frequency, system sizes and driving amplitude.
From Fig.~2, we see that varying the frequency produces
no major shift in $T_r$.
Hence we obtain
\begin{equation}
T_r^{\rm 1D}\approx J.
\label{tr1dsim}
\end{equation}
In Fig.~2, the scaled SNR is plotted for
a variety of combinations of $N$, $V$ and $A$. 
The fact that they all collapse onto one curve confirms 
the expected scaling form (\ref{snrscale}).
Its breakdown is evident only for driving amplitudes $A \agt J$, 
as shown in Fig.~3, due to nonlinear effects.
From Fig.~2 one can also observe that for temperatures below
$T_r$ the simulation data are sensitive to $A$ even for
quite small values of $A$. In this region much smaller 
$A$ than simulated are required in order to reach the asymptotic 
zero-amplitude limit. Decreasing the amplitude further, however,
would increase the statistical errors significantly.

The results in 2D are qualitatively the same for $T>T_c$ as in 1D, 
except a stronger influence of the driving frequency on
$R(T)$ is seen (see Fig.~4).
At high frequencies, a second peak gradually develops below $T_c$,
as illustrated in the magnified plot in the inset of Fig.~4.
Our simulations show that this second peak 
becomes more refined as the lattice size is increased. 
We also check the validity of the scaling relation (\ref{snrscale})
and find it fails only for fairly large driving amplitudes $A>0.2$. 
Thus, in the low-frequency, small-amplitude limit, 
the resonance peak for $T>T_c$ converges to 
\begin{equation}
T_r^{\rm 2D}\approx 1.35 \pm 0.03 T_c.
\label{tr2dsim}
\end{equation}
With that we confirm the existence of a novel characteristic
temperature distinct from $T_c$.
For higher frequencies, the peak above $T_c$ shifts slightly towards $T_c$.

For 3D  the same trend versus frequency is observed
with lower frequencies resulting in a higher $T_r$, but even in 
the low-frequency limit the sharp peak is closer to $T_c$ than in 2D: 
\begin{equation}
T_r^{\rm 3D}\approx 1.1 \pm 0.05 T_c,
\label{tr3dsim}
\end{equation}
The second peak below $T_c$ also becomes more evident 
for higher frequencies and its height is now comparable 
to that of the main peak (Fig.~5).

Turning to 4D, namely the upper critical dimension of the Ising model, 
we see clearly a twin-peak structure in $R(T)$ (Fig.~5) 
and the peak above $T_c$ is even closer to $T_c$ than in 3D:
\begin{equation}
T_r^{\rm 4D} \approx 1.05 \pm 0.05 T_c.
\label{tr4dsim}
\end{equation}

In Fig.~6 we summarize
the overall trends of the height and position of the main resonance 
at $T_r>T_c$ versus the coordination number of the lattice, $2d$,
in the low-frequency, small-amplitude limit.
The decrease of the peak height
follows roughly a power law $d^{-c}$ with $c\approx 2$.

\section{Analytical approaches and discussions}

\subsection{1D exact result}

As mentioned above, the SNR for the 1D kinetic Ising model has been
computed by Schimansky-Geier et al. \cite{schimansky}
in the low frequency limit.
For completeness, their result is recorded here
\begin{equation}
R^{\rm 1D}={\pi V A^2 \over 4 T^2} \sqrt{1-\tanh^2(\frac{2J}{T})}.
\label{schi}
\end{equation}
Comparison with simulation data is given in Fig.~5.
The agreement is excellent for small $A$, except below the peak,
where the data is more sensitive to $A$ and the $A\rightarrow 0$
limit is not reached yet. 
But the overall trend of $R$ as a function of $A$
supports the above prediction.

\subsection{mean-field theories}

For higher dimensions, no exact result on $R$ is available and
we must resort to approximations.
A naive approximation is to consider at 
high temperature that an average spin 
in different dimensions only differ
by the number of spins coupled to it, given by $2d$.
We may then make use of (\ref{schi}), derived for 1D,
to obtain an approximated formula of $R(T)$ 
for $d>1$ by replacing $J$ with an effective coupling constant 
\begin{equation}
J_e=Jd.
\end{equation}
Of course this approximation should hold only at high temperatures, but
comparison with simulations in Fig.~5 (solid lines) shows agreement
better than naively expected, especially in 2D.

Independent of the approximation, the general form of the SNR is given by
\begin{equation}
R=\frac{\pi}{2} \frac{(\Delta M)^2} { {S_0(\omega_s)}},
\label{defsnr}
\end{equation}
where ${S_0}(\omega)$ is the frequency dependent power spectrum
(noise strength),
and $\Delta M$ is the amplitude of the total magnetization induced by the
external magnetic field, i.e., the `signal' in
$M(t)=V m + \Delta M \sin(\omega_s t-\phi)$ 
where $m$ is the equilibrium magnetization per spin, and 
$\phi$ is the phase shift\cite{leung}.
Note that $Q=\pi (\Delta M)^2/2$ is just the amplitude in (\ref{spectrum}).

The simplest approach to calculate $\Delta M$ and $S_0$ 
is the mean-field (MF) approximation.
The mean-field $\Delta M$ can easily be found \cite{leung}:
\begin{equation}
(\Delta M_{\rm MF})^2=\frac{V^2 A^2}{T^2}  
(1-m_{\rm MF}^2)^2 \frac{1}{\frac{1}{\tau_{\rm MF}^2}+\omega_s^2},
\label{signal}
\end{equation}
where $m_{\rm MF}$ can be determined in the standard way 
by numerically solving the self-consistency equation 
$m_{\rm MF}=\tanh(2d J m_{\rm MF}/T)$, and
$\tau_{\rm MF}$ is the mean-field relaxation time:
\begin{equation}
\tau_{\rm MF}=\frac{1}{1-\frac{2dJ}{T}(1-m_{\rm MF}^2)}.
\label{tau}
\end{equation}
The noise strength ${S_0}^{\rm MF}(\omega)$ can be determined 
using the Wiener-Khintchin theorem:
\begin{equation}
{S_0}^{\rm MF}(\omega)= \frac{2}{\tau_{\rm MF}}
\frac{<M^2>_{\rm MF}-V^2 m_{\rm MF}^2}{\frac{1}{\tau_{\rm MF}^2}+\omega^2}.
\label{noise}
\end{equation}
The numerator can readily be found via the susceptibility to be
$(1-m_{\rm MF}^2)\tau_{\rm MF}$.
With these we obtain:
\begin{equation}
R_{\rm MF}=\frac{\pi V A^2(1-m_{\rm MF}^2)}{4 T^2} 
\label{snr_MF}
\end{equation}
We plot $R_{\rm MF}(T)$ in Fig.~5 (dashed lines).
The mean-field result agrees with simulations at high temperatures, 
but near $T_c$ it misses the peaks entirely.
Instead, the factor $1-m_{\rm MF}^2$ yields a cusp at $T_c^{\rm MF}$.
Thus, mean-field theory fails to capture the essence of
the SNR from simulations.

\subsection{high-temperature expansions}

The clue for the origin of the above discrepancy comes from re-examining
the 1D exact result.
It is instructive to rewrite $R_{\rm 1D}$ of (\ref{schi}) in a more general
form:
\begin{equation}
R^{\rm 1D}={\pi V A^2 \lambda \over 4 T} 
\label{schi2}
\end{equation}
where 
\begin{equation}
\lambda ={1\over T}\sqrt{1-\tanh^2(\frac{2J}{T})}={1\over T \cosh(2J/T)}
\label{kincoe1d}
\end{equation}
is the {\em zero-field\/} kinetic coefficient, given 
by the ratio between the susceptibility and relaxation time 
\begin{equation}
\lambda={\chi\over \tau}
\label{kincoe}
\end{equation}
for {\em general\/} dimension\cite{hh}.
For 1D, $\chi=e^{2J/T}/T$ and $\tau=1/[1-\tanh(2J/T)]$ are exact.
For higher dimensions no exact result for $\chi$, $\tau$ or $\lambda$
are known, but (\ref{schi2}) remains valid.
In fact, equation (\ref{snr_MF}) is the mean-field version of 
(\ref{schi2}), with $\lambda_{\rm MF}=(1-m_{\rm MF}^2)/T$.

In the more refined mean-field approach introduced in \cite{leung}
based on the time-dependent Ginzburg-Laudau (TDGL) equation,
the SNR can be derived and it takes the same form as (\ref{schi2}).
With that approach in 2D, $\chi T$ is accurate up to $O(v^3)$, $\tau$
up to $O(v^4)$ and hence $\lambda T=1-4v^2+O(v^4)$, where
$v=\tanh(J/T)$ is the usual high-temperature expansion (HTE) parameter.
This also yields a smooth $R(T)$ with no peak.

Of course the HTE of $\lambda T$ is available in the literature
to much higher order 
for the 2D kinetic Ising model, at least up to $O(v^{20})$\cite{hte}.
It is interesting to ask if this more accurate $\lambda$ gives
the peak in $R(T)$.  Fig.~7 plots the cumulative contributions
up to various orders $v^n$, and it is clear that even $O(v^{20})$
is not enough to yield the simulated shape of $R(T)$.
That this is the right conclusion is indicated by doing the same
procedure for 1D where the HTE is known to all orders. The result (Fig.~7)
shows the same trend as in 2D. The peak is not revealed until up 
to $n \approx 70$.
We conclude that the peak in $R(T)$ is an elusive quantity to get,
its absence is due to the inaccuracy of our approximation of
the kinetic coefficient $\lambda$ near $T_c$.
In retrospect, approximations of high-temperature nature are bound to 
fail, because their expansion parameters are close to unity near 
where the peak is supposed to be.
We must therefore address the critical region.

\subsection{critical dynamics}

From the general form (\ref{schi2}), the SNR for weak field
is proportional to the kinetic coefficient $\lambda$.
From renormalization group analysis\cite{hh},
$\tau\sim \epsilon^{-\nu z}$,
and $\chi\sim \epsilon^{-\gamma}$, where $\epsilon\propto T-T_c$, we get
$\lambda \sim \epsilon^{\nu(z-2+\eta)}$,
where $\gamma=\nu(2-\eta)$ is used.
It is then clear that $R(T)$ must exhibit a maximum near $T_c$ 
because critical slowing down entails the vanishing of the kinetic 
coefficient at $T_c$\cite{hh}, $R(T)$ must bend down near 
$T_c$ as $T$ is lowered. 

It is, however, clear from the simulations that $R(T)$ does not
really vanish at $T_c$ (finite-size effect plays no role since
the convergence with respect to $V$ already occurs at rather small $V$).
The physical reason is that the presence of an oscillating magnetic
field prevents the system from fully developing its correlations
near $T_c$ within the finite period $1/f_s$.
Singularities are rounded. In particular, critical 
slowing down is suppressed due to the cutoff of $\tau$ at $1/f_s$. 
Hence roundings are expected to occur at roughly $T^*$, where
$(T^*-T_c)/T_c \sim f_s^{1/\nu z}$.
For infinite system size, the vanishing of $R$ is then
controlled by $f_s$ according to $R(T_c)\sim f_s^{(z-2+\eta)/z}$  
($R$ is controlled by $L$ instead if $L<f_s^{-1/z}$).
Thus, for small frequencies,
we expect the kinetic coefficient to drop
significantly as $T_c$ is approached, giving rise to
the peak in the SNR, 
hence a novel temperature scale distinct from $T_c$.

Since an accurate functional form of $\lambda(T)$ near $T_c$ 
is lacking, to partially remedy the situation 
we illustrate the above idea by means of a phenomenological 
description for the SNR.
We simply replace the above `$\sim$' signs by equalities and
specify $\epsilon=(T-T_c)/T$ to get 
$\tau_{\rm RG}=\epsilon^{-\nu z}$, 
$\chi_{\rm RG} T=\epsilon^{-\nu(2-\eta)}$ and
\begin{equation}
\lambda_{\rm RG} T=\epsilon^{\nu(z-2+\eta)}.
\label{lambdaRG}
\end{equation}
For the 2D kinetic Ising model, the exponent values are $\eta=1/4$,
$\nu=1$, $z\approx 2.16$ ($z$ is not known exactly\cite{hte}).
This particular form has the obvious advantage of capturing both 
the correct high-temperature value 
(both $\tau$ and $\chi T\to 1$ as $T\to \infty$) 
and the behavior near $T_c$.  
In fact, $\tau_{\rm RG}$ and $\chi_{\rm RG}$ agree
surprisingly well with the HTE away from $T_c$.
The corresponding SNR, $R_{\rm RG}(T)$, is plotted in Fig.~8,
which indeed shows the characteristic peak at about the right place.
The scenario is qualitatively the same in higher dimensions,
but the singularities are weaker (logarithmic in 4D) 
which plausibly explains why $T_r-T_c$ decreases with $d$ (Fig.~6).

\section{conclusion}

We have simulated the kinetic Ising model in various spatial dimensions
under the influence of an oscillating magnetic field.  We focus on
the signal-to-noise ratio, $R$, as a measure of stochastic resonance
for the spins in response to the external field.
For all the dimensions we study, $R$ exhibits a clear maximum
at a resonance temperature distinct from the equilibrium critical
temperature.  Various theoretical approaches to calculate
$R$ are discussed which, when properly incorporating the 
critical slowing down at $T_c$, agree with the simulated results.

We have confined our attention to the case of weak fields, since
strong fields induce more complex response than is treatable
by linear response theories.
One such complications is the saturation of the magnetization
within one cycle of oscillation which would generate higher harmonics
in the power spectrum, thus invalidating even the usual definition of SNR.
Theoretically, we have also confined mostly to $T>T_c$.  Below $T_c$,
localized excitations such as nucleation of droplets may become 
important in certain region of the parameter space. They are more 
difficult to handle than a spatially uniform perturbation done here.

With respect to other resonance temperatures defined
by means of the correlation function between magnetization
and external field\cite{neda,leung},  the present resonance 
temperature seems to be unrelated.
Since, unlike the previous ones, 
it does not converge to $T_c$ in the small frequency limit, 
it offers a more robust characterization of
the phenomena of stochastic resonance in kinetic Ising systems.
We expect that experimental measurement of the $R(T)$ curve 
performed on mono-domain magnetic particles with 
localized magnetic moments could
reveal this novel resonance temperature.

\section{Acknowledgments}

We are grateful to the NSC of R.O.C. for their support through the grant
NSC87-2112-M-001-006 and NSC88-2112-M001-013.



\onecolumn


\begin{figure}[htp]
\epsfig{figure=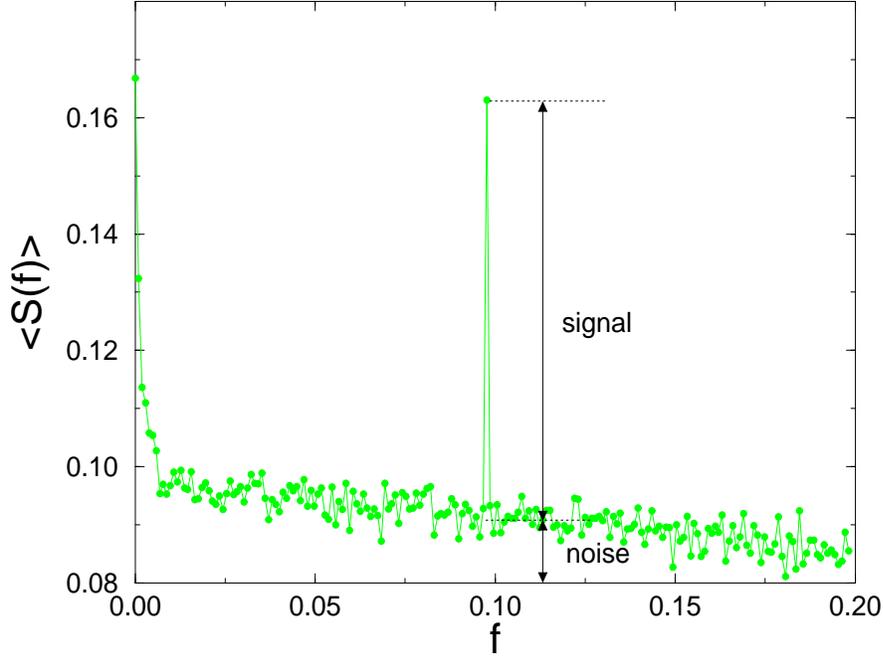,height=4.0in,width=5.0in,angle=-0}
\caption{Characteristic shape of $<S(f)>$ 
for a driving frequency $f_s\approx 0.097$.}
\label{fig1}
\end{figure}


\begin{figure}[htp]
\epsfig{figure=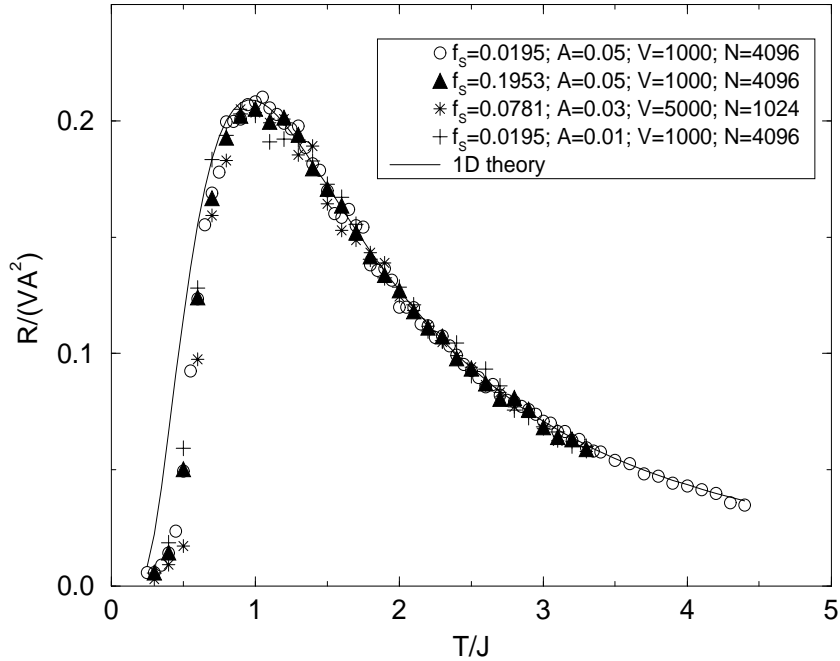,height=4in,width=5.0in,angle=-0}
\caption{General shape of the $R(T)/VA^2$ 
curve in 1D for different driving frequency $f_s$, driving amplitude $A$,
system size $V$, and time step $N$. 
The continuous line is the exact theoretical
result (\ref{schi}). }
\label{fig2}
\end{figure}


\begin{figure}[htp]
\epsfig{figure=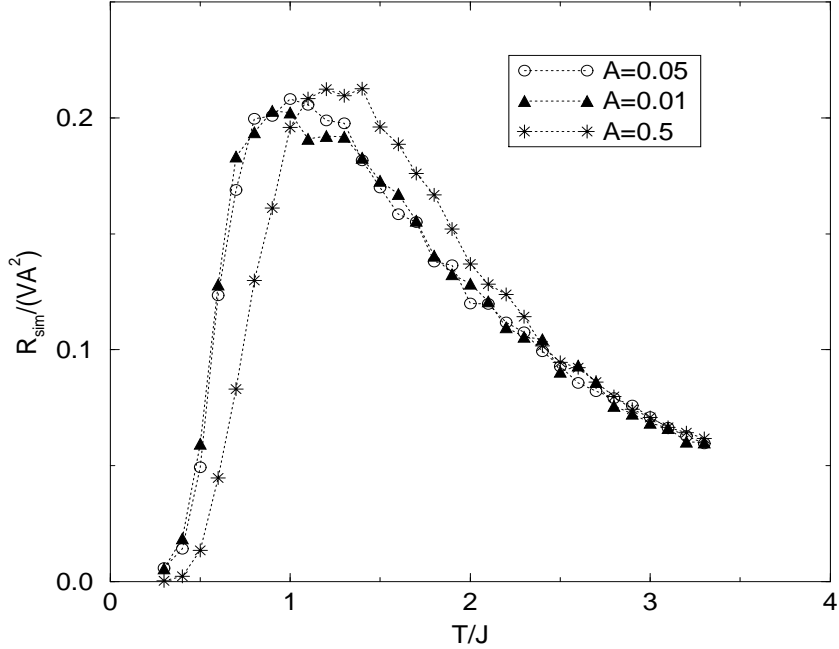,height=4.0in,width=5.0in,angle=-0}
\caption{Break-down of the scaling law  (\ref{snrscale}) 
for high driving field intensity, $A$.
$N=4096$, $V=1000$ and $f_s=0.019531$ for all curves.
}
\label{fig3}
\end{figure}


\begin{figure}[htp]
\epsfig{figure=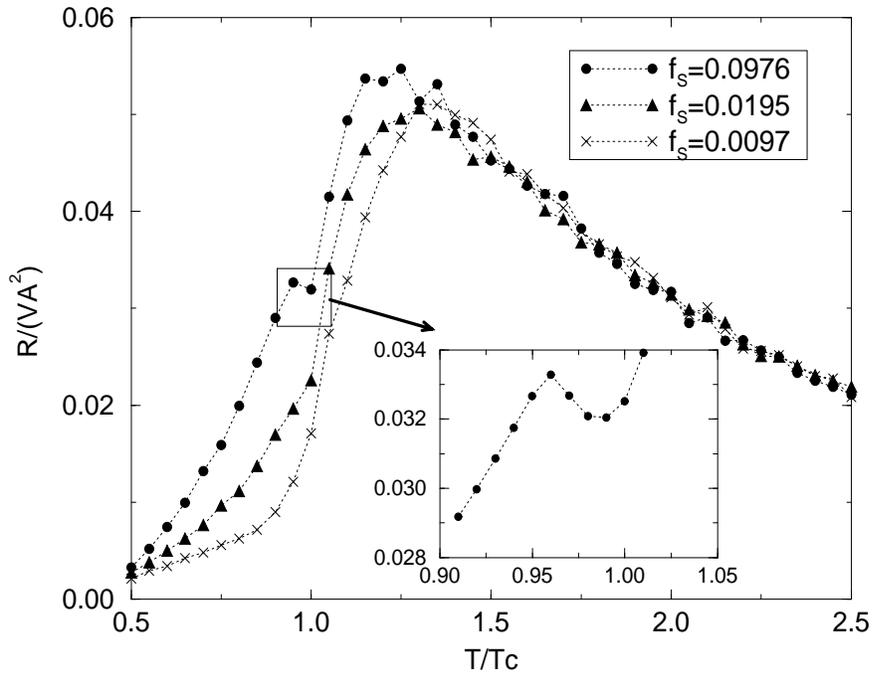,height=4in,width=5.0in,angle=-0}
\caption{General shape and frequency dependence of
the $R(T)/VA^2$ curves in 2D. All results are for $V=50\times50$,
$N=1024$ and $A=0.11$. The magnified region shows the peak under
$T_c$, observable for high driving frequencies.}
\label{fig4}
\end{figure}


\begin{figure}[htp]
\epsfig{figure=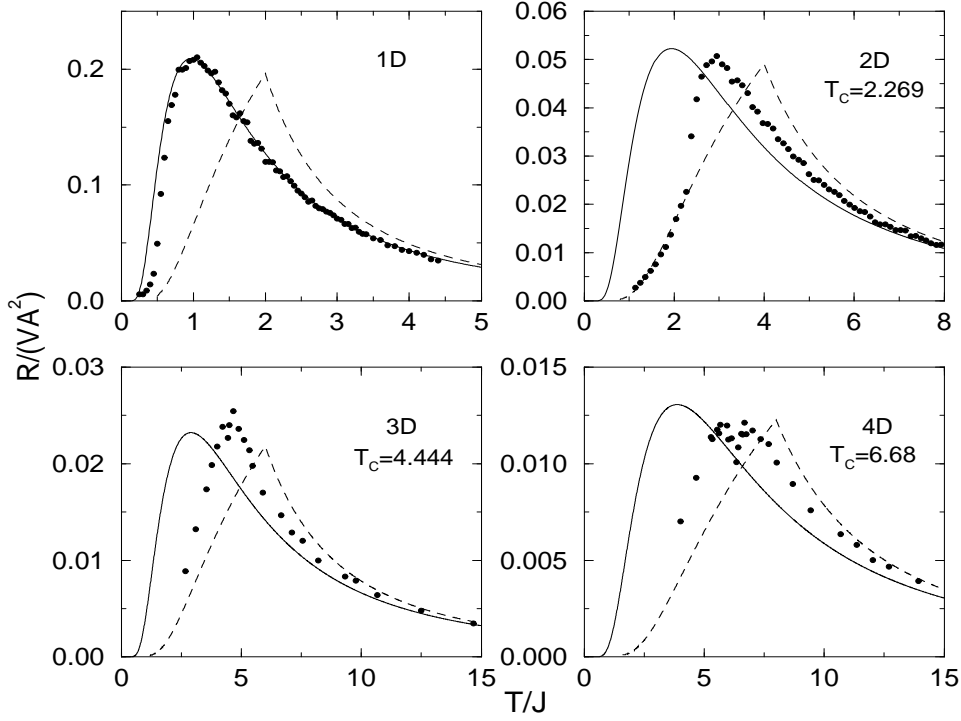,height=4.0in,width=5.0in,angle=-0}
\caption{Modified 1D theory with $J_e=Jd$  
(solid line), and MF approximation 
(dashed line) in comparison with characteristic simulation data for 
all the considered dimensions.}
\label{fig5}
\end{figure}


\begin{figure}[htp]
\epsfig{figure=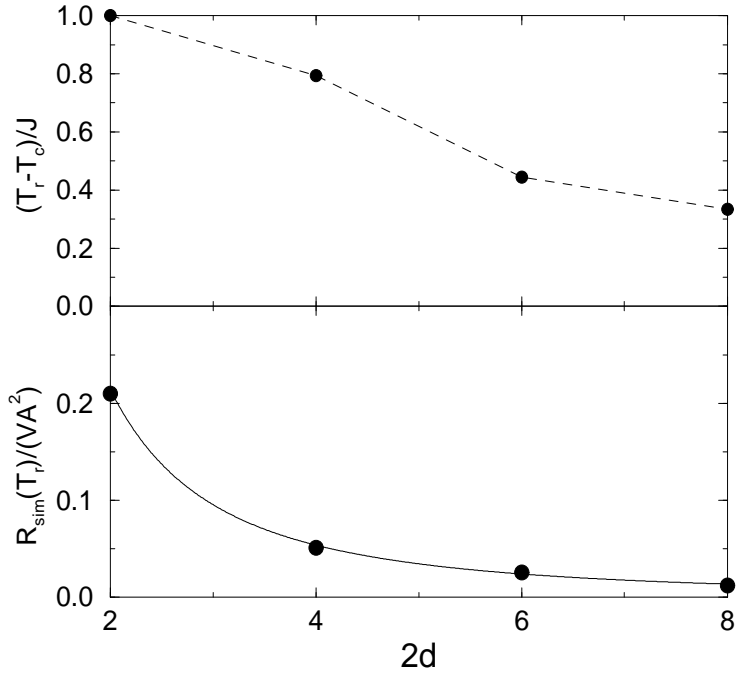,height=4in,width=5.0in,angle=-0}
\caption{Trend for $T_r$ and height of the peak as a function of 
the dimensionality of the square-type lattices. The decrease of
the height is well fitted by a $d^{-c}$ power law with $c=2$.}
\label{fig6}
\end{figure}


\begin{figure}[htp]
\epsfig{figure=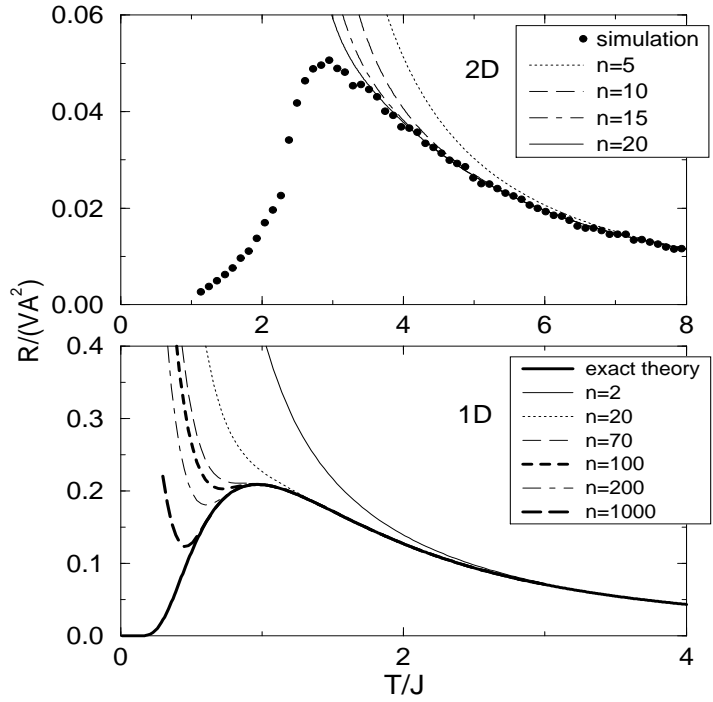,height=4.0in,width=5.0in,angle=-0}
\caption{High temperature expansion results as a function of
$v=tanh(J/T)$ and $v=tanh(2J/T)$ for 2D and 1D systems, respectively.
($n$ indicates the order of the expansion.)}
\label{fig7}
\end{figure}


\begin{figure}[htp]
\epsfig{figure=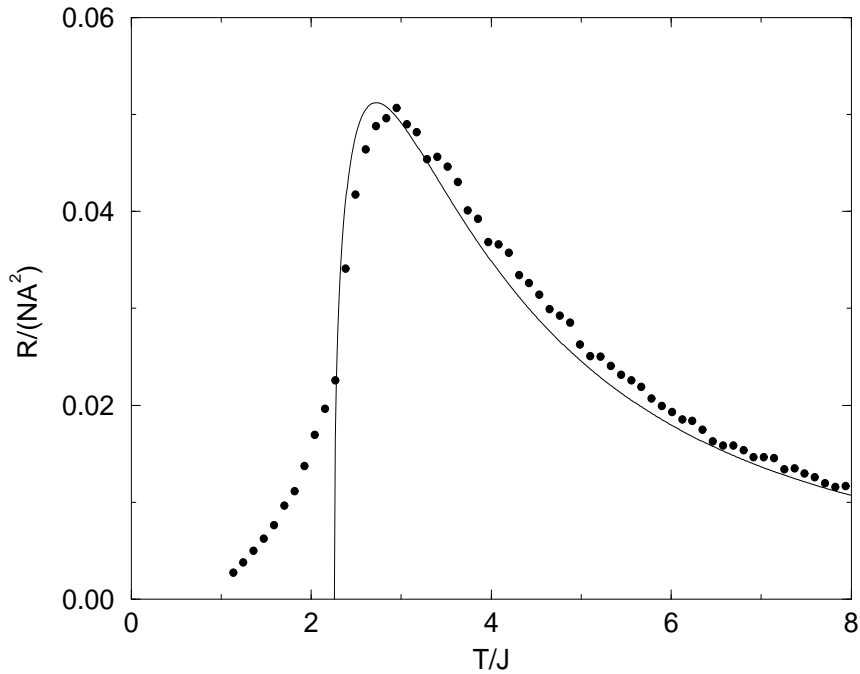,height=4in,width=5.0in,angle=-0}
\caption{Result using the phenomenological approximation
(\ref{lambdaRG}) for the kinetic coefficient (solid line)
in comparison with 2D simulation data.}
\label{fig8}
\end{figure}

\end{document}